%% file: 0paper.tex
\documentclass[letterpaper]{article} 
\usepackage{aaai2026}
\usepackage{times}  
\usepackage{helvet}  
\usepackage{courier}  
\usepackage[hyphens]{url}  
\usepackage{graphicx} 
\usepackage{natbib}  
\usepackage{caption} 
\usepackage{algorithm}
\usepackage{algorithmic}
\usepackage{xcolor}
\usepackage{amsmath}
\input{00packages}
\usepackage{newfloat}
\usepackage{listings}
\DeclareCaptionStyle{ruled}{labelfont=normalfont,labelsep=colon,strut=off} 
\floatstyle{ruled}
\newfloat{listing}{tb}{lst}{}
\floatname{listing}{Listing}
\usepackage{booktabs}

\title{Are Gains Quiet and Losses Loud? Emotional Responses to Financial Booms and Crashes Online}
\author{
    Aryan Ramchandra Kapadia\equalcontrib\textsuperscript{\rm 1},
    Niharika Bhattacharjee\equalcontrib\textsuperscript{\rm 1}, 
    Mung Yao Jia\equalcontrib\textsuperscript{\rm 1},
    Ishq Gupta\textsuperscript{\rm 1}, \\ 
    Dong Wang\textsuperscript{\rm 1},
    Koustuv Saha\textsuperscript{\rm 1}
}
\affiliations{
    \textsuperscript{\rm 1}University of Illinois Urbana-Champaign\\
    \{kapadia8, nb23, myjia2, ig8, dwang24, ksaha2\}@illinois.edu

}

\begin{document}

\maketitle 

\begin{abstract}
   \input{0abstract}
\end{abstract}
\input{1introduction}
\input{2relatedwork}
\input{3data}

\input{4methods}

\input{5rq1}
\input{6rq2}

\input{7discussion}

\fontsize{9pt}{8pt} {\selectfont
\bibliography{references}
}

\section{Ethics Statement}
In our study, we collect and analyze the language of financial and non-financial Reddit communities to explore the impact of financial booms and crashes on user emotions, sentiment, and psycholinguistic patterns using quasi-experimental frameworks. We recognize the sensitive nature of studying users’ behavior and well-being. Additionally, the classification models used in the analysis have limitations, as user expressions can be complex and may not directly match the available model labels.  Consequently, there is a risk of misinterpreting people’s expressions of emotion, sentiments, and psycholinguistic patterns. We also acknowledge that future work exploring the impact of financial-related stressors on users’ mental health requires careful consideration of external factors, such as demographics, to avoid perpetuating biases and stereotypes. As a result, building future interventions in this area would require several iterations and discussions with users before deployment can be considered. 
\appendix
\input{8appendix}

\end{document}

%% file: 00packages.tex
\usepackage{color, colortbl}
\usepackage[table]{xcolor}
\usepackage{url}
\usepackage{subcaption}
\usepackage{textcomp}
\usepackage{soul}
\usepackage{multirow}
\usepackage{enumitem}
\usepackage{mathtools}
\usepackage{siunitx}

\usepackage{booktabs} 
\usepackage{longtable,booktabs}

\usepackage{arydshln}
\definecolor{linkColor}{RGB}{6,125,233}
\definecolor{green}{rgb}{0.0, 0.65, 0.31}
\definecolor{bleudefrance}{rgb}{0.19, 0.55, 0.91}
\definecolor{ceruleanblue}{rgb}{0.16, 0.32, 0.75}
\definecolor{grey}{HTML}{969696}
\definecolor{violet}{HTML}{756bb1}
\definecolor{dgrey}{HTML}{01665e}
\definecolor{lgrey}{HTML}{5ab4ac}
\definecolor{dgreen}{HTML}{005a32}
\definecolor{purple}{HTML}{ae017e}

\definecolor{editCol}{HTML}{000000}
\definecolor{maskCol}{HTML}{c51b7d}
\definecolor{lrColor}{HTML}{8856a7}
\definecolor{trColor}{HTML}{d01c8b}
\definecolor{ctColor}{HTML}{4dac26}
\definecolor{brickred}{HTML}{f03b20}

\definecolor{sigone}{RGB}{222, 235, 247}    
\definecolor{sigtwo}{RGB}{158, 202, 225}    
\definecolor{sigthree}{RGB}{107, 174, 214}  

\definecolor{improveCol}{HTML}{4dac26}
\definecolor{worsenCol}{HTML}{d01c8b}

\definecolor{DarkBlue}{HTML}{00008B}
\definecolor{mscolor}{HTML}{01665e}
\definecolor{nmscolor}{HTML}{bf812d}
\definecolor{lgreen}{HTML}{ccece6}
\definecolor{dolive}{HTML}{308014}

\colorlet{tablerowcolor4}{gray!50} 

\newcommand*{\textlabel}[2]{%
  \edef\@currentlabel{#1}
  \phantomsection
  #1\label{#2}
}

\colorlet{tableheadcolor}{gray!25} 
\colorlet{tablerowcolor}{gray!10} 
\colorlet{tablerowcolor2}{gray!45} 
\colorlet{tablerowcolor3}{gray!12} 

\newcolumntype{a}{>{\columncolor{tablerowcolor}}r}
\definecolor{aicolor}{HTML}{018571}
\definecolor{occolor}{HTML}{ff7799}

\definecolor{aicolor}{HTML}{fc8d62}
\definecolor{occolor}{HTML}{253494}

\newif{\ifhidecomments}
  \hidecommentsfalse 
\ifhidecomments
    \newcommand{\mungyao}[1]{}
    \newcommand{\niharika}[1]{}
    \newcommand{\aryan}[1]{}
    \newcommand{\ishq}[1]{}
    \newcommand{\koustuv}[1]{}
\else
    \newcommand{\mungyao}[1]{\textbf{\small\sffamily{\textcolor{violet}{[#1 -- Mung Yao]}}}}
    \newcommand{\niharika}[1]{\textbf{\small\sffamily{\textcolor{DarkBlue}{[#1 -- Niharika]}}}}
    \newcommand{\aryan}[1]{\textbf{\small\sffamily{\textcolor{dgreen}{[#1 -- Aryan]}}}}
    \newcommand{\ishq}[1]{\textbf{\small\sffamily{\textcolor{orange}{[#1 -- Ishq]}}}}
    \newcommand{\koustuv}[1]{\textbf{\small\sffamily{\textcolor{purple}{[#1 -- Koustuv]}}}}
  \fi

\colorlet{tableheadcolor}{gray!25} 

\definecolor{neutralCol}{HTML}{dd1c77}
\definecolor{neutralGreen}{HTML}{31a354}
\definecolor{NewBlue}{HTML}{1879ba}
\definecolor{bleudefrance}{rgb}{0.19, 0.55, 0.91}  
\definecolor{AfTrColor}{HTML}{0868ac}  
\definecolor{BfTrColor}{HTML}{a8ddb5}  

\definecolor{AfCtColor}{HTML}{b10026}  
\definecolor{BfCtColor}{HTML}{fd8d3c}

\graphicspath{ {figures/} }

\newcommand{\para}[1]{\vspace{0.2em}\noindent\textbf{\textit{#1}~}}

%% file: 0abstract.tex
Financial events negatively affect emotional well-being, but large-scale studies examining their impact on online emotional expression using real-time social media data remain limited. To address this gap, we propose analyzing Reddit communities (financial and non-financial) across two case studies: a financial crash and a boom. We investigate how emotional and psycholinguistic responses differ between financial and non-financial communities, and the extent to which the type of financial event affects user behavior during the two case study periods. To examine the effect of these events on expressed language, we analyze daily sentiment, emotion, and LIWC counts using quasi-experimental methods: Difference-in-Differences (DiD) and Causal Impact analyses during a financial boom and a financial crash. Overall, we find coherent, negative shifts in emotional responses during financial crashes, but weaker, mixed responses during booms. By exploring emotional and psycholinguistic expressions during financial events, we identify future implications for understanding online users' mental health and building connected, healthy communities. 

%% file: 1introduction.tex
\section{Introduction}
Periods of economic uncertainty affect people's daily lives, influencing their financial decisions, emotional states, and patterns of expression. Prior research has linked financial stressors to elevated depression and psychological distress at the population-level~\cite{kokaliari2018quality}. At a theoretical level, behavioral finance and psychology offer well-established explanations of these effects. In particular, the \textit{loss aversion theory} postulates that individuals experience losses more intensely than equivalent gains, leading to amplified emotional reactions during market downturns~\cite{kahneman2013prospect}. Relatedly, the psychological theory of negativity bias suggests that negative events have a stronger effect on people's psychological well-being~\cite{rozin2001negativity}.

Although these theories are widely supported by laboratory experiments, surveys, and economic models, understanding how they manifest on people's well-being in everyday life, particularly in real time and at a large scale, remains challenging. Traditional methods rely heavily on retrospective self-reports or aggregated indicators, which are limited in their ability to capture immediate emotional responses, temporal dynamics, and collective sense making during unfolding financial events~\cite{tourangeau2000psychology}. In contrast, social media responses provides a naturalistic lens into how people publicly articulate emotions, seek advice, and navigate uncertainty as events unfold~\cite{saha2021life}. Moreover, responses can differ based on participation among different online social media communities~\cite{park2018examining}. During major financial or policy shocks, online discussions often intensify, offering signals of emotional, cognitive, and psychosocial responses related to money and well-being. 

Despite this opportunity, existing social media research has mostly focused on isolated financial crises, descriptive sentiment trends, or domain-specific outcomes (e.g., trading behavior), rather than systematically testing how different financial event types affect well-being and emotional responses~\cite{johnson2023cryptocurrency}. There is limited empirical evidence examining whether emotional and psycholinguistic responses to the types of financial events differ (e.g., booms vs. crashes) in magnitude, direction, and persistence~\cite{goodell2023emotions}. To address these gaps, we investigate how financial booms and crashes causally shape emotional and psycholinguistic expression in different online communities. Leveraging large-scale Reddit discussions, we compare and identify responses among finance-focused and non-financial communities to financial events. We adopt a quasi-experimental causal design to examine both immediate and sustained effects of financial shocks on online discourse. Our work is guided by the research questions (RQs) below:

\begin{description}
  \item[\textbf{RQ1:}] How do financial booms and crashes shape emotional and psycholinguistic expressions in online financial communities, compared to non-financial communities? 
  \item[\textbf{RQ2:}] Do financial booms and crashes produce sustained and asymmetric emotional responses in financial communities? 
\end{description}

We conduct an observational study on Reddit to analyze how user behavior may differ during a financial crash and a boom. We focus on Reddit to study real-time and long-form expressions in common-interest communities~\cite{proferes2021studying}. Prior work indicates that exogenous shocks, such as COVID-19, affect language and help-seeking behavior~\cite{saha2020psychosocial}. We further extend this by (a) focusing on how financial booms and crashes influence online behavior and (b) contextualizing broader mental-health patterns on how financial stress manifests in online expressions. 

For RQ1, we use Difference-in-Differences (DiD) to examine emotions and psycholinguistic expressions in a financial group relative to a non-financial group during financial events. For RQ2, we use causal impact (CI)~\cite{brodersen2017package} to explore the sustained and asymmetric effects of financial events on emotional responses. \textit{We find that the financial crash triggers stronger, more directionally aligned negative shifts in emotional expressions than weaker, ambivalent shifts during the boom.} Understanding users' well-being during financial events has important implications for (1) designing interventions to mitigate financial stress-related mental health risks, and (2) facilitating healthy online communities.
\newline
\textbf{Ethical Considerations} We collect public and anonymized data via APIs that comply with Reddit's terms of service to protect user privacy and avoid disclosing identifiable details. We do not require human studies or ethics board approval. 

%% file: 2relatedwork.tex
\section{Related Work}
\para{Psychological Theories Behind Loss and Gain.}
Psychologists study \textit{negativity bias}, which proposes that negative events may influence people's well-being more than positive events~\cite{rozin2001negativity}. Similarly, behavioral finance studies human decision-making through \textit{prospect theory}, which describes how people evaluate gains and losses asymmetrically~\cite{altman2010prospect, kahneman2013prospect}, and \textit{loss aversion theory}, which posits that losses are experienced more intensely than gains~\cite{schmidt2005loss}. While loss aversion theory has been studied for impacts on investors' financial decisions~\cite{hwang2010loss}, we aim to examine how these psychological theories influence the emotional and psycholinguistic responses of Reddit users during financial booms and crashes. 

\para{Psychological Impact of Financial Events.}
Surveys have reported that global financial crises are associated with depression, anxiety, and stress~\cite{butterworth2009financial, kokaliari2018quality}. However, surveys are challenging to capture users’ real-time emotions during crises, to scale across geographic regions, and to avoid recall bias~\cite{saha2020psychosocial}. Researchers have used social media such as Reddit to study the impact of financial losses on well-being, finding negative sentiment expressed in financial subreddits during market crashes and losses~\cite{johnson2023cryptocurrency}. Despite these insights, there is limited research that leverages real-time social media conversations to investigate psychological theories of financial behavior, such as loss aversion, during both financial booms and crashes. Addressing this gap could provide a more nuanced understanding of how individuals emotionally respond to financial risks and gains in real time, complementing survey-based approaches. 

\para{Causal Inference with Social Media Data.}
Causal inference models can be used to explore causal relationships between events and users' online behavior~\cite{de2016discovering,saha2022social,kiciman2018using}.
Prior work has examined how major real-world events shape online behavior on social media~\cite{chowdhury2021examining,saha2020psychosocial,olteanu2018hateful,de2014narco,mark2012blogs,lin2014ripple}.
Further, researchers have studied causal relations, such as the effect of positive reinforcement on Reddit post quality and frequency~\cite{lambert2025does}, the effect of Reddit bans on communities on user posting behavior~\cite{chandrasekharan2017you}, and the effect of COVID-19 on college students' online mental health expressions~\cite{saha2025mental}.

%% file: 3data.tex
\section{Data and Methods}

\subsection{Financial and Non-Financial Groups}
We identify active, well-established candidate subreddits with a US-centric user base categorized into \textbf{(a) financial} and \textbf{(b) non-financial} groups. Financial communities focus on topics like personal finance, investing, and debt (\textit{e.g., r/personalfinance}), while non-financial communities focus on general discussions not centered on finance (\textit{e.g., r/explainlikeimfive}). We initially selected \textbf{15 financial} and \textbf{18 non-financial} candidate subreddits (Table~\ref{tab:reddit_communities_unbalanced}), which form the starting pool for constructing balanced groups.

\subsection{Event Definitions}
We select two economy-wide financial events: a \textbf{boom} on \textit{November 8, 2024}~\cite{mikolajczak2024sp500}, and a \textbf{crash} on \textit{April 2, 2025}~\cite{wsj2025trumptariffs}. Event dates are anchored to public announcement timings rather than market trading dates to better capture users' exposure to the information. For each event, we define 30-day pre- and post-event windows to capture short- to medium-term impacts on online behavior. To establish pre-event comparability between the financial and non-financial groups, we define a \textbf{financially stable period} from \textit{April 6, 2023} to \textit{May 6, 2023}, free of major market shocks. This window is used to balance these groups and is not used to answer RQs.

\subsection{Reddit Data Collection \& Preprocessing}
 We collect posts, comments, and metadata (e.g., number of comments, upvotes) using the Arctic Shift tool~\cite{Heitmann2025} during the financial events. 
 We analyze both posts and comments to distinguish between individual self-disclosure and aggregate community responses. We only analyze de-duplicated, English-language, human-written textual content
 (data processing detailed in Appendix A.2).  

\subsection{Creating Balanced \& Comparable Groups}
~\label{sec:data_match}
To isolate the impact of the financial event from the subreddit activity~\cite{stuart2010matching}, we follow prior studies~\cite{saha2018social} to balance the financial and non-financial groups using pre-event engagement metrics that measure community activity. These include daily post volume, average comments per post, median upvotes per post, daily users, and proportion of textual posts (described in Appendix~\ref{sec:appendix_preprocessing}). 

We use \textit{bidirectional pruning} to iteratively compute standardized mean differences (SMDs) across covariates and remove subreddits that contribute most to the imbalance until all covariates fall below the maximum acceptable SMD threshold of 0.25~\cite{rubin2001propensity}. The final balanced set filtered to \textbf{15 financial} and \textbf{11 non-financial} subreddits (listed in Table~\ref{tab:reddit_communities}) with a maximum SMD of 0.166 (shown in Figure~\ref{fig:loveplot}). In our study, we analyze $\sim$320K posts and $\sim$2.8M comments, summarized in Table~\ref{tab:data_amount} and described in Table~\ref{tab:summary_stats}. To prevent cross-group contamination, we exclude users who were active in both groups during the financial boom (5,246) and the crash (30,206).

\begin{table}[htbp]
\centering
\resizebox{\columnwidth}{!}{
\begin{tabular}{p{1.3cm}p{11cm}}
\textbf{Grouping} & \textbf{Subreddit Communities}\\
\midrule
Financial &
r/personalfinance, r/investing, r/bogleheads, r/stocks, r/tax,
r/financialindependence, r/economy, r/economics,
r/debtfree, r/financialplanning, r/frugal, r/insurance,
r/careerguidance, r/creditcards, r/stockmarket\\
\midrule
Non-Financial &
r/explainlikeimfive, r/trueoffmychest, r/casualconversation,
r/askmen, r/tooafraidtoask, r/changemyview, r/asksocialscience,
r/askhistorians, r/askacademia, r/parenting, r/seriousconversation\\
\bottomrule
\end{tabular}}
\caption{List of subreddit communities in balanced groups}
\label{tab:reddit_communities}
\end{table}

\begin{table}[htbp]
\centering
\resizebox{\columnwidth}{!}{
\begin{tabular}{lccccc}
 & \multicolumn{2}{c}{\textit{Financial}} &
   \multicolumn{2}{c}{\textit{Non-Financial}} \\
\cmidrule(lr){2-3} \cmidrule(lr){4-5}
\textbf{Event} &
\textbf{Period} & \textbf{\#} &
\textbf{Period} & \textbf{\#} \\
\midrule
\multicolumn{5}{l}{\textit{\textbf{Posts}}} \\
\textit{Boom} &
08-Oct-24--08-Dec-24 & 64,473 &
08-Oct-24--08-Dec-24 & 66,269 & \\
\textit{Crash} &
02-Mar-25--02-May-25 & 83,551 &
02-Mar-25--02-May-25 & 60,943 & \\
\textit{Stability} &
06-Apr-23--06-May-23 & 24,510 &
06-Apr-23--06-May-23 & 20,940 & \\
\textbf{Overall} &
 & \textbf{172,534} &
& \textbf{148,152} & \\
\midrule
\multicolumn{5}{l}{\textit{\textbf{Comments}}} \\
\textit{Boom} &
08-Oct-24--08-Dec-24 & 397,415 &
08-Oct-24--08-Dec-24 & 893,878 & \\
\textit{Crash} &
02-Mar-25--02-May-25 & 726,553 &
02-Mar-25--02-May-25 & 831,102 & \\
\textbf{Overall} &
 & \textbf{1,123,968} &
& \textbf{1,724,980} & \\
\bottomrule
\end{tabular}}
\caption{Post and Comment counts in balanced groups}
\label{tab:data_amount}
\end{table}

%% file: 4methods.tex
\subsection{Conducting Language Analysis} 
\label{sec:emotion_sentiment_LIWC} We labeled the emotion, sentiment, and psycholinguistic attributes in posts and comments to study user language and well-being. Positive, neutral, or negative \textit{sentiment} was labeled using VADER, a sentiment analysis model~\cite{hutto2014vader}. 
\textit{Emotions} were labeled as anger, disgust, fear, joy, neutral, sadness, or surprise using a fine-tuned  DistilRoBERTa model~\cite{hartmann2022emotionenglish}. We analyzed \textit{LIWC} linguistic features~\cite{tausczik2010psychological}. We used \textit{affect} (positive emotion, negative emotion, sadness, anger, and anxiety) to measure emotional well-being, \textit{motives} (reward and risk) to measure the sense of losses and gains, and  \textit{psychological processes} (cognitive processing, tentativeness, certainty, and achievement) to measure confidence and motivation. For analysis, we calculate the daily proportion of posts containing an expression per group (financial vs. non-financial), accounting for the day$\times$subreddit-level. This normalizes posting frequency across subreddits within each group, assigning equal weights to each subreddit.

%% file: 5rq1.tex
\section{RQ1: Emotions and Psycholinguistic Expressions During Financial Events}
\label{sec:rq1}
\subsubsection{Difference-in-Differences (DiD)}
We adopted the difference-in-differences (DiD) framework to analyze the causal effect of the financial event on emotional responses in financial groups relative to non-financial groups, using Equation~\ref{eq:equation1} (detailed in Table~\ref{tab:equation_explanation}). $\text{Fin}_i$ is a group indicator (financial vs non-financial). If pre-event outcomes are parallel, any post-event shift can be attributed to the financial event (detailed in Appendix~\ref{sec:did_assumptions}). The coefficient $\beta_3$ isolates changes in outcomes ($Y_{it}$) in the financial group relative to the non-financial group due to the financial event.    
\begin{equation}
\label{eq:equation1}
Y_{it} = \alpha + \beta_1 Fin_i + \beta_2 Event_t + \beta_3(Fin_i \times  Event_t) + \beta_4 t + \varepsilon_{it}
\end{equation}

We run DiD for a (i) 30-day and (ii) 10-day pre- and post-event to capture medium- and short-term effects, respectively. DiD's run separately for posts (individual behavior) and comments (community responses) to avoid correlated errors, since comments are directly related to posts.

\subsection{DiD Short-Term Results}
During the \textit{financial crash}, the financial group's posts increased by 3.63 percentage points (pp) in surprise ($p < 0.001$) and by 3.39 pp in certain language ($p = 0.029$), relative to the non-financial group. The financial group's fear score increased by 0.67 pp relative to the non-financial group ($p = 0.013$). During the \textit{financial boom}, the financial group’s posts decreased 4.25 pp in reward-related ($p = 0.006$) and 2.06 pp in anger-related language ($p = 0.048$), while comments decreased 0.87 pp in anger ($p=0.022$), increased 1.32 pp in joy ($p = 0.028$) and 1.74 pp in cognitive processing ($p = 0.003$) relative to the non-financial group.

\begin{table}[htbp]
\centering
\label{tab:short_term_effects}
\resizebox{\columnwidth}{!}{
\begin{tabular}{lllllcc}
\textbf{Method} &\textbf{Event} & \textbf{Level} & \textbf{Expression} & \textbf{Direction} & $\boldsymbol{\beta_3}$ & \textbf{p-value} \\
\midrule
DiD & Crash & Posts & Surprise (Emotion) & $\uparrow$ & 0.0363 & $<$0.001 \\
DiD & Crash & Posts & Certainty (LIWC) & $\uparrow$ & 0.0339 & 0.029 \\
DiD & Crash & Comments & Fear (Emotion) & $\uparrow$ & 0.0067 & 0.013 \\
\midrule
DiD & Boom & Posts & Reward (LIWC) & $\downarrow$ & -0.0425 & 0.006 \\
DiD & Boom & Posts & Anger (LIWC) & $\downarrow$ & -0.0206 & 0.048 \\
DiD & Boom & Comments & Joy (Emotion) & $\uparrow$ & 0.0132 & 0.028 \\
DiD & Boom & Comments & Anger (Emotion) & $\downarrow$ & -0.0087 & 0.022 \\
DiD & Boom & Comments & Cognitive Processing (LIWC) & $\uparrow$ & 0.0174 & 0.003 \\
\bottomrule
\end{tabular}}
\caption{Short-term ($\pm$10 days) effects of financial events}
\label{tab:short_term_effects}
\end{table}

\subsection{DiD Medium-Term Results}
During the \textit{financial crash}, the financial group's \textit{posts} decreased by 2.22 pp in positive sentiment ($p=0.045$), decreased by 2.50 pp in achievement-related language ($p=0.030$), and increased by 2.34 pp in negative sentiment ($p=0.016$) relative to the non-financial group. The financial group's \textit{comments} increased by 0.98 pp in anger ($p = 0.019$) and by 0.97 pp in sadness ($p = 0.005$) relative to the non-financial group. During the \textit{financial boom}, the financial group's \textit{comments} increased by 0.64 pp in surprise ($p=0.038$), 1.09 pp in joy ($p=0.038$), 1.74 pp in positive sentiment ($p = 0.005$),  1.34 pp in reward-related language ($p = 0.027$), 1.40 pp in achievement-related language ($p = 0.021$), and by 1.42 pp in certain language ($p = 0.020$) relative to the non-financial group.

\begin{table}[htbp]
\centering
\resizebox{\columnwidth}{!}{
\begin{tabular}{lllllcc}
\textbf{Method} & \textbf{Event} & \textbf{Level} & \textbf{Expression} & \textbf{Dir.} & \textbf{Effect} & \textbf{Significance} \\
\midrule
DiD & Crash & Posts & Positive Sentiment & $\downarrow$ & $-0.0222$ & $p=0.045$ \\
DiD & Crash & Posts & Negative Sentiment & $\uparrow$ & $0.0234$ & $p=0.016$ \\
DiD & Crash & Posts & Achievement (LIWC) & $\downarrow$ & $-0.0250$ & $p=0.030$ \\
DiD & Crash & Comments & Anger (Emotion) & $\uparrow$ & $0.0098$ & $p=0.019$ \\
DiD & Crash & Comments & Sadness (Emotion) & $\uparrow$ & $0.0097$ & $p=0.005$ \\
\midrule
CI & Crash & Posts & Tentativeness (LIWC) & $\downarrow$ & $-2.09\%$ & P.P.=99.6\% \\
CI & Crash & Comments & Positive Sentiment & $\downarrow$ & $-2.54\%$ & P.P.=100\% \\
CI & Crash & Comments & Negative Sentiment & $\uparrow$ & $+6.49\%$ & P.P.=100\% \\
\midrule
DiD & Boom & Comments & Surprise (Emotion) & $\uparrow$ & $0.0064$ & $p=0.038$ \\
DiD & Boom & Comments & Joy (Emotion) & $\uparrow$ & $0.0109$ & $p=0.038$ \\
DiD & Boom & Comments & Positive Sentiment & $\uparrow$ & $0.0174$ & $p=0.005$ \\
DiD & Boom & Comments & Reward (LIWC) & $\uparrow$ & $0.0134$ & $p=0.027$ \\
DiD & Boom & Comments & Achievement (LIWC) & $\uparrow$ & $0.0140$ & $p=0.021$ \\
DiD & Boom & Comments & Certainty (LIWC) & $\uparrow$ & $0.0142$ & $p=0.020$ \\
\midrule
CI & Boom & Posts & Sadness (LIWC) & $\downarrow$ & $-10.11\%$ & P.P.=100\% \\
CI & Boom & Posts & Anger (LIWC) & $\uparrow$ & $+18.73\%$ & P.P.=100\% \\
CI & Boom & Comments & Negative Sentiment & $\uparrow$ & $+4.67\%$ & P.P.=100\% \\
CI & Boom & Comments & Joy (Emotion) & $\downarrow$ & $-7.87\%$ & P.P.=100\% \\
CI & Boom & Comments & Affect (LIWC) & $\uparrow$ & $+1.53\%$ & P.P.=100\% \\
CI & Boom & Comments & Positive Emotion (LIWC) & $\uparrow$ & $+1.62\%$ & P.P.=100\% \\
\bottomrule
\end{tabular}}
\caption{Medium-term ($\pm$ 30 days) effects of financial events. DiD reports $\beta_3$; causal impact reports relative effects.}
\label{tab:long_term_effects}
\end{table}

%% file: 6rq2.tex
\section{RQ2: Sustained Causal Effects of Financial Event on Online Financial Communities }
\label{sec:rq2}
\subsubsection{Causal Impact}
We examine whether financial crashes and booms produce asymmetric, persistent changes in emotional responses within financial communities, focusing on deviations from expected behavior absent a financial event. Comparing the magnitude and persistence of these effects allows us to assess whether crashes exert a stronger influence than booms, consistent with negativity bias and loss aversion. We estimate these deviations using a Bayesian Structural Time Series model over a 30-day window. For each sentiment, emotion, and LIWC category, we model daily averaged proportions in financial subreddits, using corresponding non-financial subreddit series as covariates. The model is trained on 30 days of pre-event data to predict a counterfactual post-event trajectory, and the causal effect is defined as the difference between the observed and predicted values. Statistical significance is defined as posterior probability $> 95\%$ and the 95$\%$ credible interval excluding zero.

\subsection{Causal Impact Results}
For the \textit{financial boom}, sadness-related language decreased by 10.11\% (P.P.\% = 100\%), but anger-related language increased by 18.73\% (P.P.\% = 100\%) in posts. For the \textit{financial crash}, tentative language decreased by 2.09\% (P.P.\% = 99.6\%), suggesting increased decisiveness in posts during the crisis. \textit{The comments have prominent sustained effects.} For the \textit{financial crash}, negative sentiment increased by 6.49\% (P.P.\% = 100\%), while positive emotion decreased by 2.54 \% (P.P.\% = 100\%). For the \textit{financial boom}, negative sentiment increased by 4.67 \% (P.P.\% = 100\%), and joy decreased by 7.87\% (P.P.\% = 100\%) in comments. Affective language increased by 1.53\% (P.P.\% = 100\%) and positive emotion language by 1.62\% (P.P.\% = 100\%).

%% file: 7discussion.tex
\section{Discussion}
The \textit{DiD analysis} reveals a clear asymmetry in how financial, relative to non-financial, communities react to losses versus gains (Tables~\ref{tab:short_term_effects} and~\ref{tab:long_term_effects}). During the crash, the financial group exhibits a unidirectional, negative affect in expressed language, compared to the non-financial group. These include medium-term increases in negative sentiment, anger, and sadness, alongside decreased positive sentiment and achievement-oriented language. \textit{We interpret these shifts as signals of increased distress and negative emotional responses during periods of financial strain.} These effects appear in both posts and comments, but medium-term effects are more prominent in comments, suggesting that \textit{community-level interactions may sustain and amplify the negative emotional responses}. In contrast, these outcomes are weaker and more heterogeneous during the financial boom. Specifically, comments show moderate increases in positive sentiment, surprise, joy, achievement, certainty, and reward-oriented language. In contrast, post-level effects are short-lived, characterized by a conflicted reduction in reward-oriented language. Hence, \textit{these inconsistent signals do not reflect clear or sustained shifts in mental well-being during the boom.}

The \textit{causal impact analysis} also supports this asymmetric pattern (Table~\ref{tab:long_term_effects}). During the crash, community responses show fewer but more coherent negative effects, including increases in negative sentiment and decreases in positive sentiment. Financial boom produces mixed impacts: we observe a decrease in sadness, but an increase in anger for the posts, and an increase in negative sentiment, a decrease in joy, but an increase in affective and positive language in comments. This incongruity suggests that crashes have a more pronounced negative impact on affect than booms at the community level. Taken together, \textit{our findings suggest that financial crashes serve as stronger and more persistent stressors in online communities than financial booms in fostering positive engagement}. This is consistent with the asymmetric responses predicted through the behavioral theory of loss aversion.

\section{Conclusion}
To conclude, we analyze emotional and psycholinguistic responses to a financial boom and crash across matched financial (15) and non-financial (11) groups using quasi-experimental methods. Both analyses reveal a common pattern: \textit{crashes trigger stronger, more directionally aligned negative shifts in emotional and psycholinguistic expressions compared to weaker, ambivalent shifts due to booms}. Heterogeneous boom signals suggest that positive financial events do not uniformly translate to collective positivity. \textit{While posts capture short-term individual affect, comments significantly sustain and amplify these responses, especially during the crash.} Our results align with empirical findings of past studies on loss aversion~\cite{schmidt2005loss} and contribute to the use of social media data to explore users' real-time responses to financial events. 

Beyond financial decision-making, our findings complement prior research on how high-stakes negative events influence online behavior, such as the impact of the COVID-19 pandemic on college students' mental health expressions~\cite{saha2025mental}. The pronounced negative shifts in emotional responses during the crash align with prior work on the psychological effects of financial stress~\cite{sturgeon2016psychosocial} and motivate similar mental-health studies. By incorporating psycholinguistic patterns in response to financial stressors, our findings can augment prior frameworks of understanding people's mental health, such as identifying suicidal ideation~\cite{shimgekar2025detecting,de2016discovering,yuan2023mental}, depression~\cite{de2013predicting}, or stress levels~\cite{guntuku2019understanding,saha2017stress} during financial crashes. Building on prior social support research~\cite{de2014mental, yang2019seekers}, these insights can also inform event-aware interventions and platform designs that incorporate resource recommendations (e.g., support links during financial crises) to reduce stress-related mental health risks.

\para{Limitations and Future Directions.}
Our work has limitations, which also suggest interesting future directions. Our work focuses on US-centric events and Reddit; cultural and platform-specific differences may affect responses across geographies and social-media platforms, which we leave for future work. While we present results from two notable case studies, we recognize that they may not generalize to all financial events. Rather, we provide a framework for studying online emotional responses to financial events, which can be extended to future case studies. Although we follow a quasi-experimental design, we cannot claim ``true'' causality because counterfactual responses cannot be observed. Additionally, both frameworks depend on stable pre-event relationships between the financial and non-financial groups and may under-capture short-term, volatile effects, especially due to anticipatory market behavior and aggregation choices. In the future, these limitations can be addressed by explicitly modeling such behavior and exploring alternate, higher-resolution counterfactual constructions.

%% file: 8appendix.tex
\section{Appendix}
\renewcommand{\thetable}{\thesection\arabic{table}}
\renewcommand{\thefigure}{A\arabic{figure}}
\setcounter{table}{0}
\setcounter{figure}{0}
\subsection{Data Details}
\label{sec:data_dets}
We summarize the Reddit metadata that we collect in Table~\ref{tab:variables}. 

\begin{table}[htbp]
  \resizebox{\columnwidth}{!}{
    \begin{tabular}{ll}
    Reddit Variable &  Description\\
    \midrule
    id & Post and comment id\\
    subreddit & The name of the subreddit\\ 
    author & Reddit username of the post author\\ 
    created\_utc & Time of creation of the post\\ 
    title & The title of the post\\ 
    selftext & The text content of the post\\ 
    num\_comments & Number of comments on the post\\ 
    ups & Upvotes received on the post\\ 
    upvote\_ratio & Upvotes/Total Number of Votes\\
    subreddit\_subscribers & Number of Subscribers on the Subreddit \\
    \bottomrule
  \end{tabular}}
  \caption{Summary of the raw data variables that we will use for our analysis.}
  \label{tab:variables}
\end{table}

\subsection{Data Preprocessing Details }
We detail our data preprocessing steps below. 
\label{sec:appendix_preprocessing}
\subsubsection{Content Exclusions} 
We filtered the raw data and excluded posts and comments that are irrelevant to conversational analysis:
\begin{itemize}
    \item NSFW, spoilers, locked or archived posts by using the flags: \texttt{over\_18}, \texttt{spoiler}, \texttt{locked}, and \texttt{archived}.
    \item Poll posts by using the flag, \texttt{poll\_data}.
    \item Deleted/removed posts and comments, identified using the markers in the title and text for posts and the body for comments ("[removed]", "[deleted]"). 
    \item We also exclude comments shorter than 10 characters to exclude low-information responses such as \textit{lol, thanks, okay}.
\end{itemize}

\subsubsection{Automoderator and bot content} We excluded posts and comments created by \texttt{Automoderator} and bot accounts, which we identified using conservative bot heuristics (username ends in \_bot or -bot, or names that contain \texttt{bot} as a unique token). 

\subsubsection{URL Handling} 
We extracted and counted all URLs from the \texttt{selftext} field in posts and the \texttt{body} field in comments. The following features were generated, which may be useful in our analysis: \texttt{has\_url}, \texttt{count\_url}, \texttt{url\_domains}. We then replaced the URLs in the text with the token \texttt{<URL>}, while preserving the Markdown text. This helped us reduce lexical noise while retaining semantic context.

\subsubsection{Removing the duplicates} We removed duplicate posts based on the same \texttt{id}, or a combination of \texttt{title, text, selftext}. This helped us eliminate duplicate posts without losing the revised versions. Additionally, we also remove duplicate comments based on the \texttt{id} field.

\subsubsection{Time Normalization}
We converted the data timestamps to Eastern Time (America/New York). It is crucial to align on ET time because major US indices and market events are timestamped in ET. We use ET in our data analysis. 

\subsubsection{Language filter} 
We used the English-only posts and comments for our analysis. 
The raw data does not have any language markers. These were generated using the \texttt{langdetect} module in Python.

\subsubsection{Text Normalization}
We normalized \texttt{author}, \texttt{title}, and \texttt{selftext} to lowercase and Unicode, and normalized spacing. We created \texttt{clean\_text} by combining the canonized \texttt{title} and \texttt{selftext}. We use the same normalization process on the \texttt{body} field of the comments to create the \texttt{clean\_text} field, which is used in the downstream analysis.

\subsubsection{Removing Overlapping Users} For the financial crash, we excluded 1,186 users for posts and 29,020 users for comments. For the financial boom, we exclude 940 users for posts and 4,306 users for comments.

\subsection{Matching Reddit Communities}
\subsubsection{Engagement Metrics Details}
To match Reddit communities to create two comparable financial and non-financial groups we collect the following variables: calendar day average posts per day (measures how active a subreddit is daily), average number of comments per post (measures the mean magnitude of discussion of a post), median upvotes per post (captures the typical community approval of a post while being robust to extremely viral posts), average number of unique users per day (reflects the size of the active user base in the community), and has\_text (indicates the percentage of posts that have textual content rather than only links or media; this allows our analysis to focus on text-based outcome variables). 

\subsubsection{Reddit Communities in Unbalanced Financial and Non-Financial Groups}
We include a list of subreddits included in the financial and non-financial groups before the balancing procedure in Table~\ref{tab:reddit_communities_unbalanced}.
\begin{table}[htbp]
\centering
\resizebox{\columnwidth}{!}{
\begin{tabular}{lcl}
\toprule
\textbf{Grouping} & \textbf{Count} & \textbf{Subreddit Communities} \\
\midrule
Financial & 15 &
r/personalfinance, r/investing, r/bogleheads, r/stocks, \\
& & r/financialindependence, r/economy, r/economics, r/tax, \\
& & r/debtfree, r/financialplanning, r/frugal, r/insurance, \\
& & r/careerguidance, r/creditcards, r/stockmarket \\
\midrule
Non-Financial & 18 &
r/explainlikeimfive, r/trueoffmychest, r/casualconversation, r/askmen, \\
& & r/tooafraidtoask, r/changemyview, r/asksocialscience, r/askhistorians, \\
& & r/askacademia, r/parenting, r/seriousconversation, r/advice, \\
& & r/askdocs, r/legaladvice, r/nostupidquestions, \\
& & r/relationship\_advice, r/offmychest, r/relationships \\
\bottomrule
\end{tabular}}
\caption{List of subreddit communities in unbalanced financial and non-financial groups.}
\label{tab:reddit_communities_unbalanced}
\end{table}

\begin{figure}[htbp]
  \centering
  \includegraphics[width=\columnwidth]{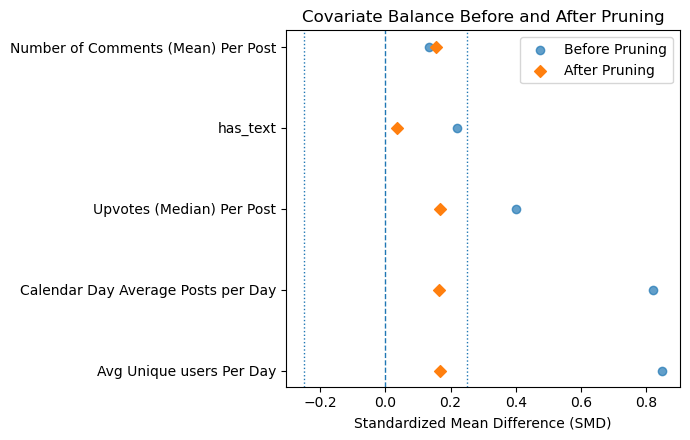}
  \caption{Love plot of Covariate Balance (SMD)}
  \label{fig:loveplot}
\end{figure}

\subsection{Descriptive Statistics for Reddit Posts}
We include a summary of statistics for the Reddit posts we collected across the financial stable, boom, and crash periods in Table~\ref{tab:summary_stats}. 

\begin{table}[htbp]
\centering
\resizebox{\columnwidth}{!}{
\begin{tabular}{llcc}
\textbf{Period} & \textbf{Metric} & \textbf{Financial} & \textbf{Non-Financial} \\
\midrule
\multicolumn{4}{l}{\textit{Financially Stable}} \\
& Calendar Day Average Posts per Day & 765.938 & 654.375 \\
& Number of Comments (Mean) Per Post & 18.386 & 30.225 \\
& Upvotes (Median) Per Post & 1.000 & 2.000 \\
& Avg Unique Users Per Day & 744.323 & 648.226 \\
& has\_text & 0.873 & 0.740 \\
& Average\_sentiment & 0.387 & 0.740 \\
& Total Number of Posts & 24,510 & 20,940 \\
& Total Number of Posts with Text & 21,552 & 15,620 \\
& Number of Subreddits & 15 & 11 \\
\midrule
\multicolumn{4}{l}{\textit{Financial Boom}} \\
& Calendar Day Average Posts per Day & 1,039.887 & 1,068.855 \\
& Number of Comments (Mean) Per Post & 14.359 & 30.183 \\
& Upvotes (Median) Per Post & 1.000 & 1.000 \\
& Avg Unique Users Per Day & 999.967 & 1,021.656 \\
& has\_text & 0.916 & 0.856 \\
& Average\_sentiment & 0.398 & 0.151 \\
& Total Number of Posts & 64,473 & 66,269 \\
& Total Number of Posts with Text & 59,125 & 56,789 \\
& Number of Subreddits & 15 & 11 \\
\midrule
\multicolumn{4}{l}{\textit{Financial Crash}} \\
& Calendar Day Average Posts per Day & 1,347.597 & 982.952 \\
& Number of Comments (Mean) Per Post & 21.606 & 30.483 \\
& Upvotes (Median) Per Post & 1.000 & 1.000 \\
& Avg Unique Users Per Day & 1,257.590 & 953.082 \\
& has\_text & 0.876 & 0.863 \\
& Average\_sentiment & 0.326 & 0.155 \\
& Total Number of Posts & 83,551 & 60,943 \\
& Total Number of Posts with Text & 73,215 & 52,653 \\
& Number of Subreddits & 15 & 11 \\
\bottomrule
\end{tabular}}
\caption{Summary of statistics for collected data during financially stable, boom, and crash periods for the financial and non-financial groups.}
\label{tab:summary_stats}
\end{table}

\subsection{Descriptive Statistics for Reddit Comments}
We include a summary of statistics for the Reddit comments that we collected across the financial boom and crash periods in Table~\ref{tab:summary_stats_comments}.

\begin{table}[t]
\centering
\label{tab:appendix_descriptives_comments}
\resizebox{\columnwidth}{!}{
\begin{tabular}{lcc}
\textbf{Metric} & \textbf{Boom} & \textbf{Crash} \\
\midrule
Comments (Total) & 1{,}291{,}293 & 1{,}557{,}655 \\
\quad Financial Group & 397{,}415 & 726{,}553 \\
\quad Non-Financial Group & 893{,}878 & 831{,}102 \\
\midrule
Posts with $\geq$1 Comment (\%) & 79.8 & 79.2 \\
Posts with 0 Comments (\%) & 20.2 & 20.8 \\
\midrule
Comments per Post: Mean & 12.37 & 13.48 \\
Comments per Post: Median & 4.00 & 4.00 \\
\midrule
Comments per Post (Financial Group): Mean / Median & 7.64 / 3.00 & 10.68 / 3.00 \\
Comments per Post (Non-Financial Group): Mean / Median & 17.07 / 5.00 & 17.47 / 5.00 \\
\midrule
Comment Length (words): Mean & 58.6 & 53.8 \\
Comment Length (words): Median & 35.0 & 31.0 \\
\midrule
VADER Sentiment: Mean & 0.189 & 0.138 \\
VADER Sentiment: Median & 0.202 & 0.064 \\
\bottomrule
\end{tabular}
}
\caption{Descriptive statistics for Reddit comments during the financial boom and crash periods. We report statistics for financial and non-financial communities after preprocessing.}
\label{tab:summary_stats_comments}
\end{table}

\subsection{Difference-in-Differences}
\subsubsection{DiD Equation Definitions} We define the variables in our DiD framework in Table~\ref{tab:equation_explanation}.

\begin{table}[htbp]
\centering
\resizebox{\columnwidth}{!}{
\begin{tabular}{lp{0.7\linewidth}}
\textbf{Variable} & \textbf{Variable Definition} \\
\midrule
$Y_{it}$ &
Daily outcome for group $i$ on day $t$ \\

$\alpha$ &
Baseline outcome for the non-financial group prior to the financial event \\

$\text{Fin}_i$ & Indicator equal to 1 for the financial group and 0 for the non-financial group \\

$\beta_1$ &
Baseline difference between the financial and non-financial groups before the financial event \\

$\text{Event}_t$ &
Indicator equal to 1 on and after the financial event (boom/crash) date \\

$\beta_2$ &
Post-event (boom/crash) change for the non-financial group \\

$\text{Fin}_i \times \text{Event}_t$ &
Interaction term, equal to 1 for observations in the financial group on and after the event date, and 0 otherwise \\ 

$\beta_3$ &
Coefficient of the interaction term, which captures the additional change in $Y_{it}$ that the financial group experienced after a financial event
relative to the non-financial group \\

$t$ &
Linear time trend controlling for smooth temporal dynamics \\

$\beta_4$ &
Average daily trend in the outcome variable \\

$\varepsilon_{it}$ &
Error term; standard errors are HAC-robust \\
\bottomrule
\end{tabular}}
\caption{Summary of DiD model variables and coefficients}
\label{tab:equation_explanation}
\end{table}

\subsubsection{DiD Assumptions}
\label{sec:did_assumptions}
\begin{figure*}[htbp]
  \centering
  \includegraphics[width=\textwidth]{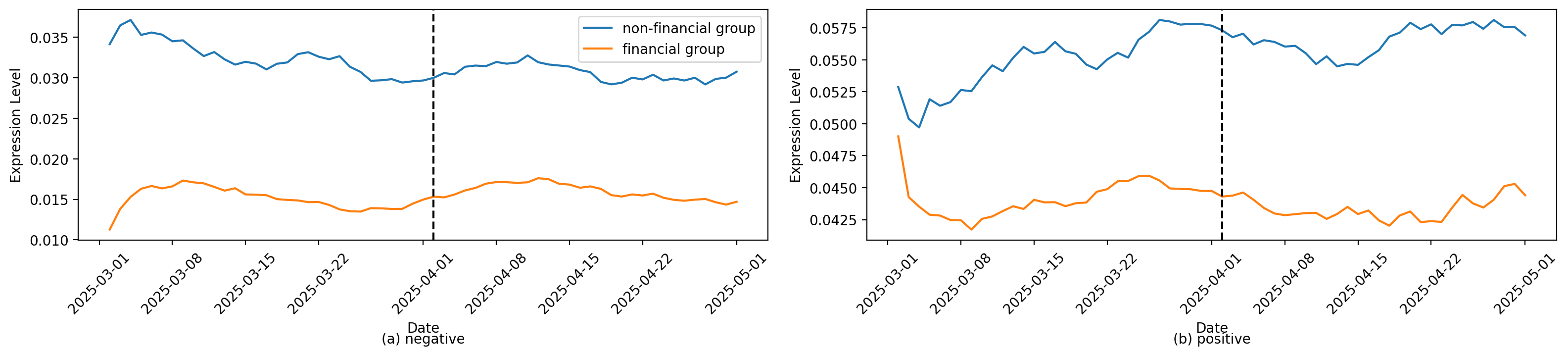}
  \caption{Daily sentiment levels (7-day rolling average) in posts during the financial crash}
  \label{fig:crash_posts_sentiment}
\end{figure*}

\begin{figure*}[htbp]
  \centering
  \includegraphics[width=\textwidth]{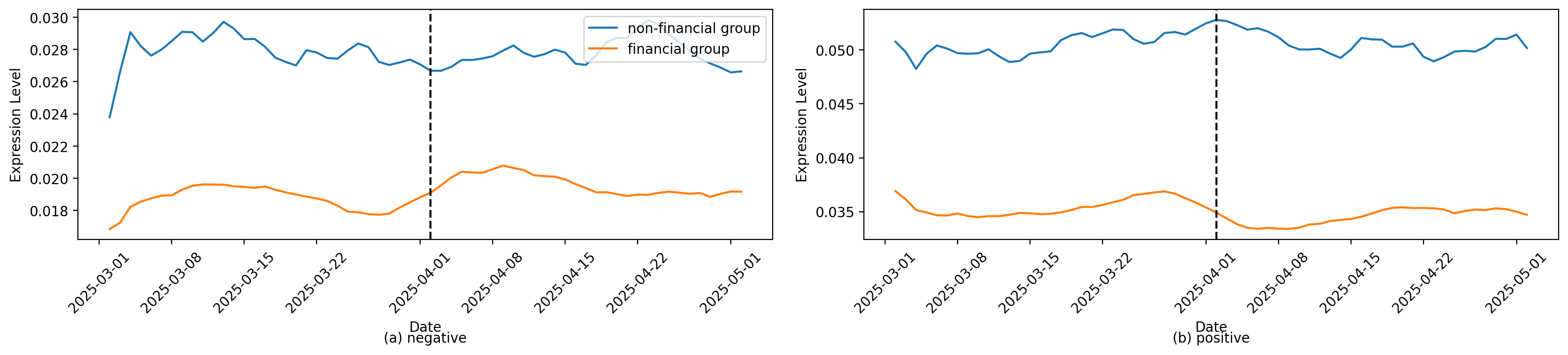}
  \caption{Daily sentiment levels (7-day rolling average) in comments during the financial crash}
  \label{fig:crash_comments_sentiment}
\end{figure*}

\begin{figure*}[htbp]
  \centering
  \includegraphics[width=\textwidth]{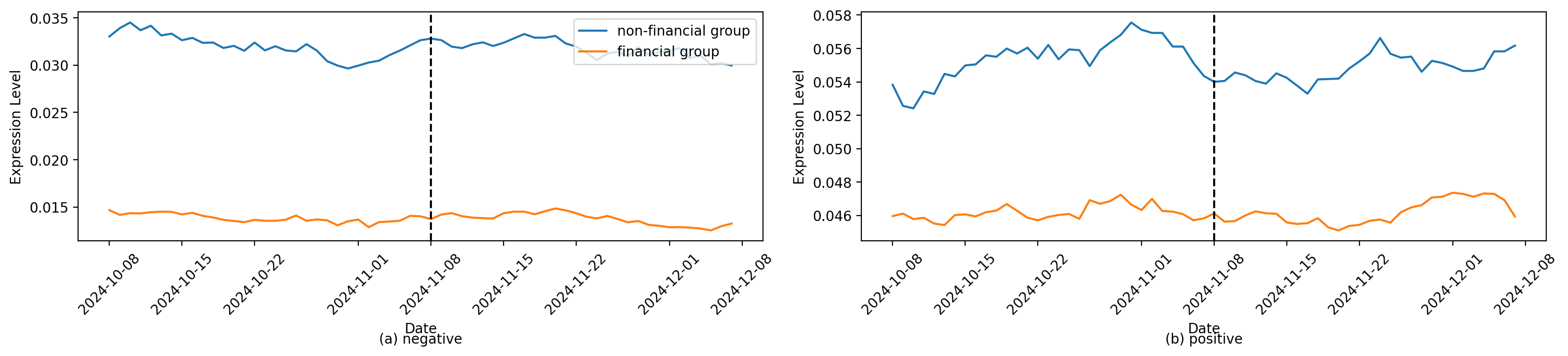}
  \caption{Daily sentiment levels (7-day rolling average) in posts during the financial boom}
  \label{fig:boom_posts_sentiment}
\end{figure*}

\begin{figure*}[htbp]
  \centering
  \includegraphics[width=\textwidth]{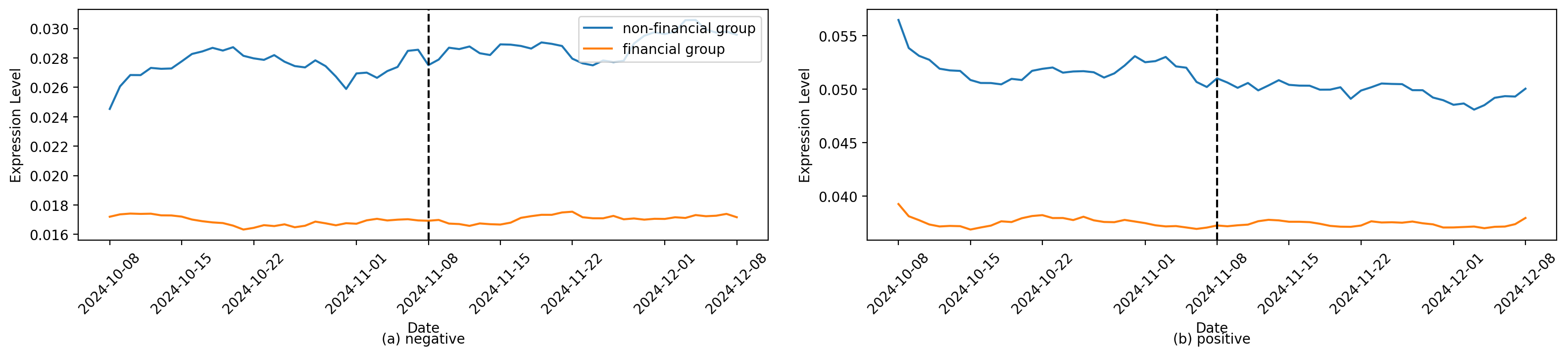}
  \caption{Daily sentiment levels (7-day rolling average) in comments during the financial boom}
  \label{fig:boom_comments_sentiment}
\end{figure*}

The primary assumption for the validity of DiD results is that the trends between the non-financial and financial groups should be parallel in the pre-event period.  As an example, we present the graph trends (7-day rolling averages) for positive and negative sentiments for the financial crash for posts in Figure~\ref{fig:crash_posts_sentiment} and for comments in Figure~\ref{fig:crash_comments_sentiment}. We present example sentiment trends for the financial boom for posts in Figure~\ref{fig:boom_posts_sentiment} and for comments inFigure~\ref{fig:boom_comments_sentiment}. We rely on visual inspection and observe a fairly parallel trend between the financial and non-financial groups for both positive and negative sentiments in posts and comments across both events. This provides reasonable support to the DiD assumption.

